\pdfoutput=1

\documentclass[aps,prx,twocolumn,showpacs,groupedaddress,notitlepage,superscriptaddress,floatfix,longbibliography]{revtex4-1}
\usepackage{amsmath,amsthm,amssymb,amsfonts,enumitem,float,graphicx,physics,hyperref}
\usepackage{color}
\usepackage[dvipsnames]{xcolor}
\usepackage[normalem]{ulem}
\usepackage{subfigure}

\newcommand{\fid}{\mathcal{F}_\mathrm{XEB}}

\theoremstyle{definition}

\newcommand{\beq}{\begin{equation}}
\newcommand{\eeq}{\end{equation}}
\newcommand{\bes} {\begin{subequations}}
\newcommand{\ees} {\end{subequations}}

\hypersetup{
    colorlinks=true, 
    linkcolor=cyan,
    citecolor=magenta, 
    filecolor=magenta, 
    urlcolor=cyan,
    runcolor=cyan
}

\begin{document}

\title{Boundaries of quantum supremacy via random circuit sampling}

\author{Alexander Zlokapa}
\affiliation{Division of Physics, Mathematics and Astronomy, Caltech, Pasadena, CA 91125, USA}
\author{Sergio Boixo}
\affiliation{Google AI Quantum, Venice, CA 90291, USA}
\author{Daniel Lidar}
\affiliation{Departments of Electrical and Computer Engineering, Chemistry, and Physics \& Astronomy, and Center for Quantum Information Science \& Technology, University of Southern California, Los Angeles, CA 90089, USA\\
PMA Moore Scholar \& INQNET Associate, Caltech, Pasadena, CA 91125, USA}

\begin{abstract}
Google's recent quantum supremacy experiment heralded a transition point where quantum computing performed a computational task, random circuit sampling, that is beyond the practical reach of modern supercomputers. We examine the constraints of the observed quantum runtime advantage in an extrapolation to circuits with a larger number of qubits and gates. 
Due to the exponential decrease of the experimental fidelity with the number of qubits and gates, we demonstrate for current fidelities a theoretical classical runtime advantage for circuits deeper than a few hundred gates, while quantum runtimes for cross-entropy benchmarking limit the region of a quantum advantage to a few hundred qubits. However, the quantum runtime advantage boundary in circuit width and depth grows exponentially with respect to reduced error rates, and our work highlights the importance of continued progress along this line. Extrapolations of measured error rates suggest that the limiting circuit size for which a computationally feasible quantum runtime advantage in cross-entropy benchmarking can be achieved approximately coincides with expectations for early implementations of the surface code and other quantum error correction methods. Thus the boundaries of quantum supremacy via random circuit sampling may fortuitously coincide with the advent of scalable, error corrected quantum computing in the near term.
\end{abstract}

\maketitle

\section{Introduction}%
A recent seminal result~\cite{google} by Google Quantum AI and collaborators claimed quantum supremacy~\cite{Boixo:2016aa,Aaronson:2016aa,Preskill:2012aa,Harrow:2017aa,neill2018blueprint,Bouland:2018aa,Movassagh:2019aa}, sampling pseudo-random quantum circuits on noisy intermediate-scale quantum (NISQ) hardware~\cite{preskill2018quantum} beyond what can be done in practice by state-of-the-art supercomputers. 
The observed runtime advantage over classical simulation methods in random circuit sampling provides a critical achievement in establishing a performance benchmark between NISQ computers and classical supercomputers.   
It shows a quantum runtime speedup beyond the reach of implemented classical algorithms~\cite{Boixo:2016aa,pednault_breaking_2017,de_raedt_massively_2018,markov2018quantum,chen2018classical,Chen_2018,villalonga2019flexible,Villalonga_2020,gray2020hyper,huang2020classical}. However, it is inevitable that implemented classical algorithms will continue to improve, and there also exist proposals~\cite{ibm,zhou2020limits} for faster classical implementations which have not been realized so far. 
Moreover, as the quantum circuit width and depth increase, the fidelity of the output from Google's Sycamore quantum chip is observed to decrease exponentially.  
Thus, in the absence of quantum error correction (QEC)~\cite{Lidar-Brun:book}, an exponential number of samples in circuit size is required to perform cross-entropy benchmarking through random circuit sampling. Yet, at the same time NISQ hardware continues to improve in terms of gate and measurement fidelities, qubit coherence, and other critical parameters. In light of this race between existing and prospective quantum and classical capabilities, here we seek to quantify the longevity of the observed quantum runtime advantage in cross-entropy benchmarking with respect to quantum circuit width and depth, and gate fidelity. More broadly, we aim to place the significance of the observed advantage in the context of current and future milestones for quantum computing in the NISQ era. This era is expected to last up to the point where QEC becomes pervasive and beneficial -- the QEC era -- estimated to be around a few thousand qubits for gate-model quantum computers, assuming a concurrent improvement in metrics of fidelity and coherence.

Our approach is to combine the empirical cross-entropy benchmarking results from Ref.~\cite{google} with the known scaling of state of the art classical algorithm for random circuit sampling, to arrive at estimates for the circuit width and depth at which classical computers catch up and surpass the time taken to acquire enough samples on the Sycamore chip to solve the cross-entropy benchmarking problem (which we define precisely below). We also consider extrapolations to higher fidelities, as expected from the current progression of improvement in NISQ hardware.
Our main finding is that the quantum advantage demonstration for the cross-entropy benchmarking problem at partial fidelity is in fact a relatively limited region in the circuit width and depth plane, bounded by classical algorithms and intractable quantum runtimes. More specifically, we demonstrate for current fidelities a theoretical classical runtime advantage for circuits of depth greater than a few hundred gates, while quantum runtimes for cross-entropy benchmarking limit the region of a quantum advantage to a few hundred qubits. Improving the fidelity enlarges the area of this region, which is maintained chiefly by classical memory constraints.

Our results suggest that the boundary of advantage in quantum random circuit sampling will soon coincide with circuit sizes sufficiently large for early implementations of QEC. Anticipating the continued development of quantum devices, more significant benefits than are possible from random circuit sampling are anticipated from scaling advantages of quantum algorithms for important problems such as prime factorization~\cite{Shor:97}, matrix inversion~\cite{Harrow:2009aa} or quantum simulation~\cite{Feynman1,Lloyd:96,Lidar:98RC,Aspuru-Guzik:05,wecker2015solving,Babbush:2017aa,Reiher:2017aa,Berry:2018aa,Jiang:2018aa,babbush2018encoding}, whose operation will be ensured by QEC. Thus, our results suggest that we may witness a smooth transition from beyond-classical computation in the NISQ era to applications in the QEC era.

The structure of this paper is as follows. In Sec.~\ref{sec:RCS} we briefly review random circuit sampling and the linear cross-entropy benchmarking metric $\fid$. In Sec.~\ref{sec:errmod} we explain how we arrive at an empirical fidelity formula that allows us to fit the measured values of $\fid$ in terms of both circuit depth and width. The computational problem associated with cross-entropy benchmarking is defined in Sec.~\ref{sec:XEB}. We then combine all these ingredients and assess the expected performance of classical algorithms. We start with the Schr\"odinger algorithm in Sec.~\ref{sec:SA}, which we identify as having an asymptotic scaling advantage above a threshold depth. Our main results about the boundaries of a quantum advantage are reported in Sec.~\ref{sec:q-adv}, where we also consider the Schr\"odinger-Feynmann algorithm and tensor networks. We discuss the implications of our results in detail in Sec.~\ref{sec:implications}, and offer concluding remarks in Sec.~\ref{sec:conc}.

\section{Random circuit sampling}%
\label{sec:RCS}

In \emph{circuit sampling}, we seek to sample from the probability distribution of outcomes $p_U(x) = |\bra{x}U\ket{0}|^2$ for a given quantum circuit $U$ and bitstrings $\ket{x}$, starting from the all-zero string $\ket{0}$. In \emph{random circuit sampling} (RCS) as experimentally realized by the Google experiment~\cite{Boixo:2016aa}, we consider circuits $U \in \mathcal{U}$, defining the set of circuits $\mathcal{U}$ to be $n$-qubit circuits with $m$ cycles, where each cycle consists of a layer of randomly chosen single-qubit gates applied to all qubits followed by a layer of two-qubit gates. In reality, ideal unitary circuits will be replaced by noisy versions, generally completely-positive maps $\mathcal{E}_U$. Thus the sampling is from the noisy probability distribution of outcomes $\tilde{p}_U(x)$ given by $\text{Tr}[\ket{x}\!\bra{x}\mathcal{E}_U(\ket{0}\!\bra{0})]$. 

If the circuits are perfectly implemented, the distribution over measurement probabilities of bitstrings is Porter-Thomas (exponential), in which case RCS is classically hard for the right circuit architecture~\cite{Bouland:2018aa}. At the other extreme, if the circuits are completely noisy, the distribution is approximately uniform, in which case RCS is classically trivial. To quantify this, we use an estimator of fidelity called linear cross-entropy benchmarking (XEB) $\fid$~\cite{Boixo:2016aa,neill2018blueprint,google}. It is given by averaging over the simulated probabilities of the measured bitstrings for many random circuits $U$
\begin{align}
\label{eq:fid}
\fid = 2^n \langle p_U(x)\rangle-1 ,
\end{align}
where $p_U(x)$ is the probability of bitstring $x$ computed classically for the ideal quantum circuit, and the average is over the observed bitstrings (drawn from $\tilde{p}_U(x)$) and random circuits $U$. $\fid$ compares how often each bitstring is observed experimentally with its corresponding ideal probability computed via classical simulation.  
It can also be understood as a test that checks that the observed samples tend to concentrate on the outputs that have higher probabilities
under the ideal (Porter-Thomas) distribution for the given quantum circuit, or simply as
the probability that no error has occurred while running the circuit.
For a random, perfectly implemented quantum circuit, we have $\fid = 1$. At the other extreme, sampling from the uniform distribution will give $\fid = 0$. 

These random quantum circuits were implemented in the Google experiment without QEC and were observed to have $\fid$ large enough to distinguish the observed distribution over measurement probabilities of bitstrings from uniformly random sampling, even against classical spoofing attempts, under certain plausible conjectures~\cite{Boixo:2016aa,google,aaronson2019classical}.

\section{Empirical Fidelity Model}%
\label{sec:errmod}

\begin{figure}[t]
\centering
\includegraphics[width=0.5\textwidth]{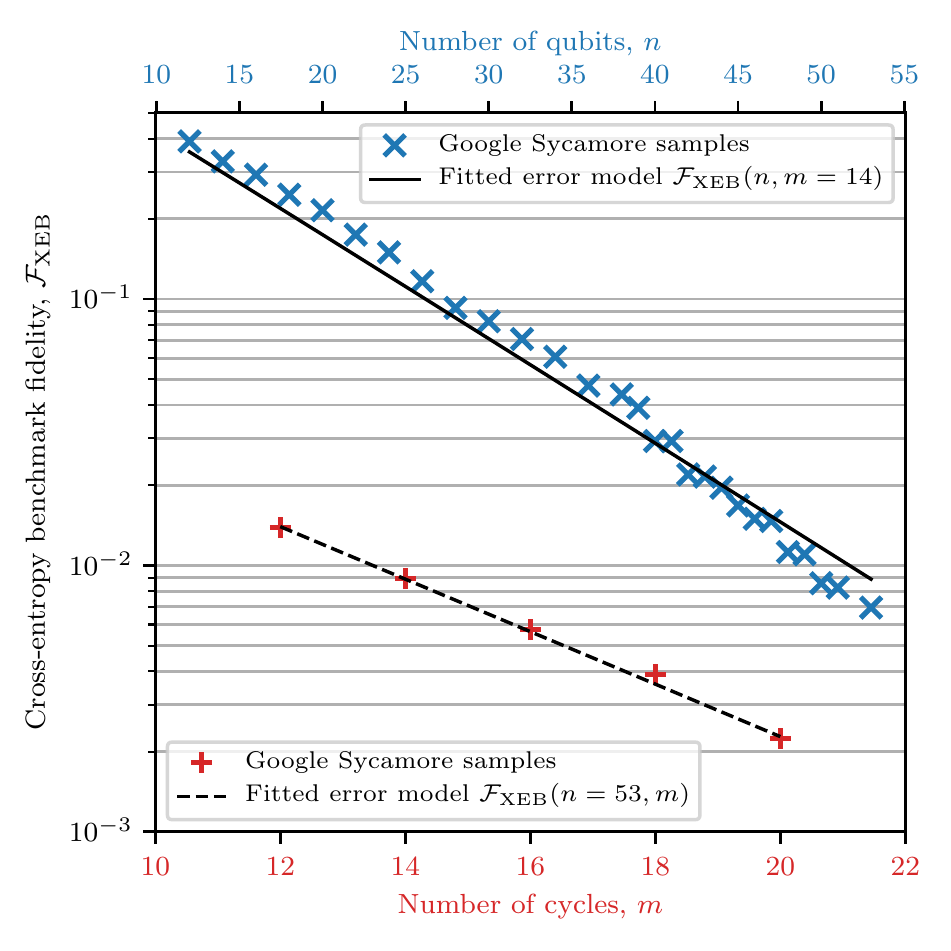}
\caption{(color online) Empirical fidelity model Eq.~\eqref{eq:F} for $\lambda = 0.0043 \pm 0.0008$ and $\gamma = 0.042 \pm 0.017$ (to two standard deviations) showing quality of fit for elided verifiable circuits of fixed depth (blue) and elided supremacy circuits of fixed width (red). The deviation from the fit is largely caused by the proportion of two-qubit gates used in larger circuits. Source: Fig.~4 of Ref.~\cite{google}.}
\label{fig:qs}
\end{figure}

From the depolarization error model in Eq.~(77) of Ref.~\cite{supp}, we have the following fidelity:
    $F = \prod_{g\in G}(1 - e_g) \prod_{q\in Q}(1 - e_q)$
for gate set $G$, gate errors $e_g$, qubit set $Q$, and qubit errors (measurement and state preparation) $e_q$. To establish scaling without QEC, we simplify the approximation of fidelity into cycle errors [scaling as a function of $m(n + (n-\sqrt{n})/2)$, since each of $m$ cycles requires $n$ single-qubit gates and $(n-\sqrt{n})/2$ two-qubit gates due to the planar architecture tiling pattern] and qubit errors (scaling as a function of $n$ due to readout error for each qubit). We thus approximate the fidelity as 
\beq
\label{eq:F}
F = 2^{-\lambda m(3n-\sqrt{n})/2 - \gamma n}, 
\eeq
and perform an empirical fit to data from Ref.~\cite{google} shown in Fig.~\ref{fig:qs} since the cross-entropy benchmark fidelity $\fid$ is 
a good estimator of the above fidelity $F$~\cite{Boixo:2016aa,markov2018quantum,neill2018blueprint,google}. The parameters $\lambda, \gamma$ are constants that result from the regular application of single- and two-qubit gates, respectively, as random gate selection in RCS allows us to adopt an effective error rate per cycle or qubit. Caveats regarding the extrapolation of this error model beyond the experimentally tested circuit width and depth are discussed below.

\begin{figure*}[t]
\centering
\subfigure[\ SA]{\includegraphics[width=.95\columnwidth]{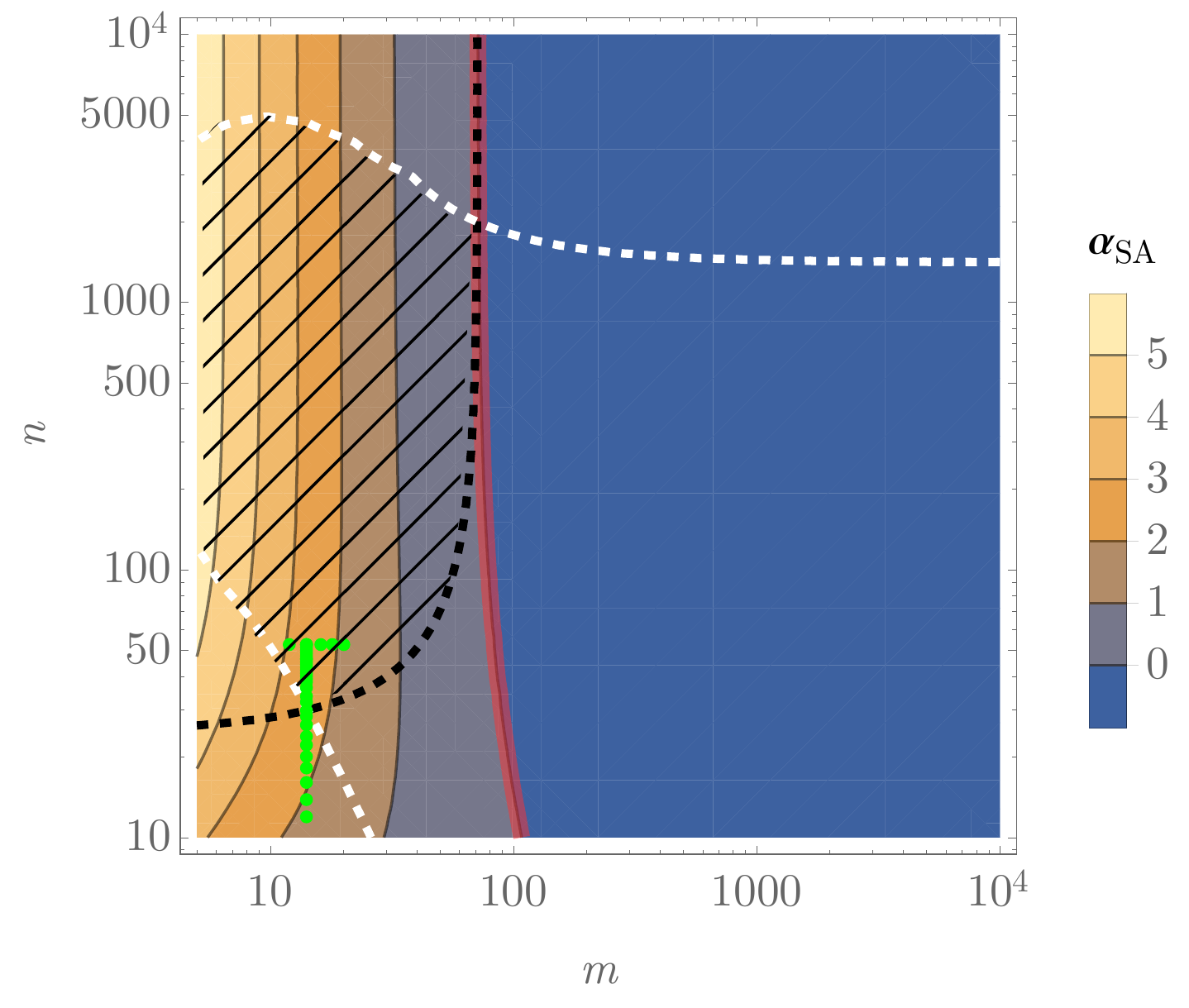}}
\subfigure[\ SFA]{\includegraphics[width=.95\columnwidth]{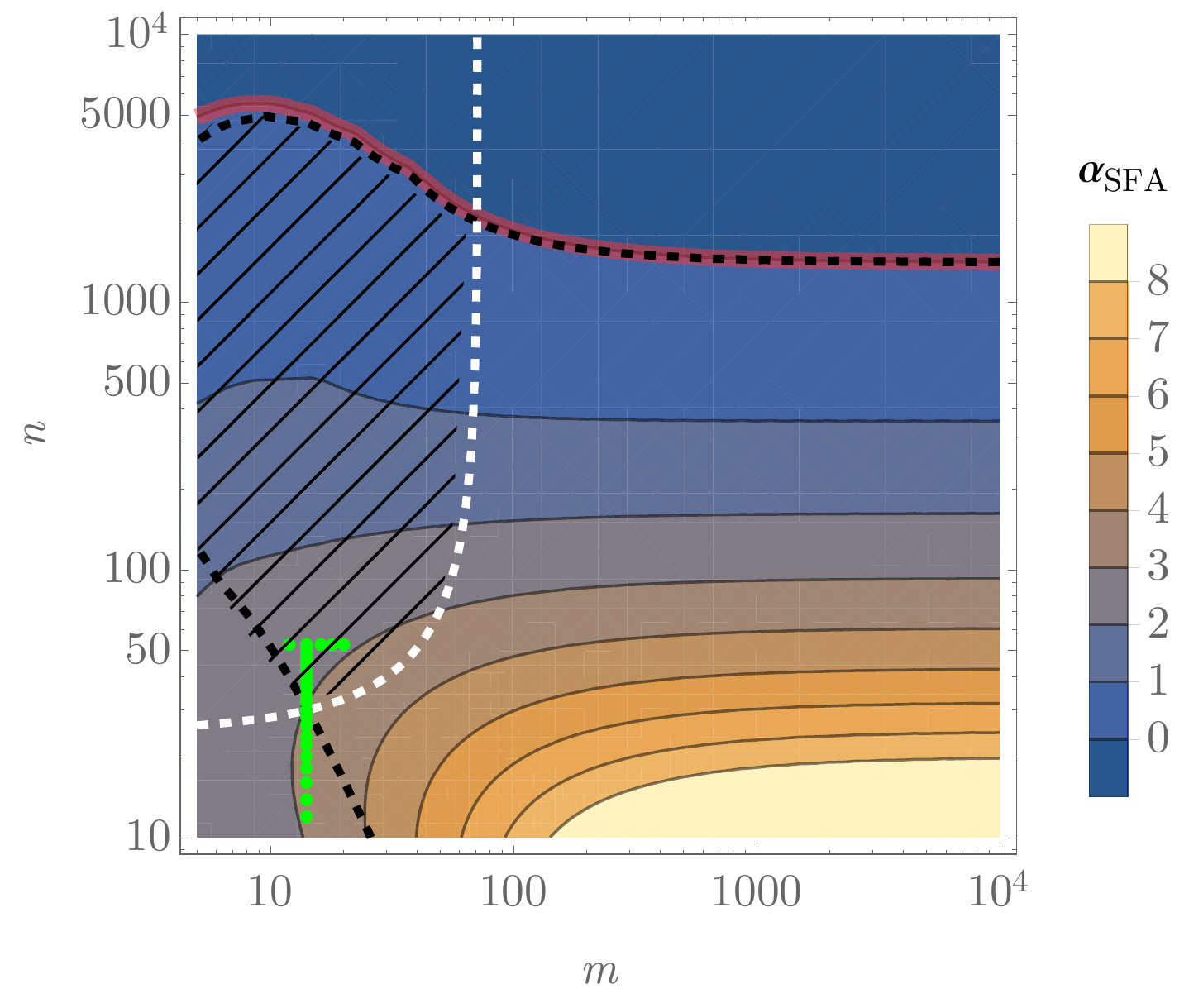}}
\caption{(color online) Effective polynomial speedup $\alpha_\text{C}$ for (a) SA and (b) SFA for circuits with $n$ qubits and $m$ cycles (taking $p \geq 2$ patches). Thick red line shows $\alpha_\text{C} = 0$, while black dashed lines indicate quantum runtime advantage boundaries $R_\text{C} = R_\text{Q}$ given by the projected \emph{runtimes} (including clock speed coefficients from Eqs. (110-111) of Ref.~\cite{supp}) of each algorithm. For SA, the quantum advantage lies to the left of the black dashed line; for SFA, the quantum advantage lies to the right of the lower black dashed line and below the upper one. The white boundaries show the bound imposed by the other algorithm, yielding a quantum runtime advantage against both SA and SFA in the region enclosed by the the black and white dashed lines (hatched). The Google Sycamore experiments are shown in green dots, entering the region of a quantum runtime advantage. Only depths larger than $5$ cycles are shown due to recent polynomial-time simulation results for shallow 2D circuits~\cite{napp2019efficient}.}
\label{fig:run}
\end{figure*}

\section{Computational problem}%
\label{sec:XEB}

We consider {\sc Cross Entropy Benchmarking} ({\sl XEB}): the problem of generating a sample of bitstrings such that the random variable $\fid(n,m)$ [Eq.~\eqref{eq:fid}] for circuits with $n$ qubits and $m$ cycles can be estimated to within standard deviation $\sigma \leq \fid$. This is the computational problem solved in Google's quantum supremacy work~\cite{google}. 

For a quantum computer (QC) the task is to take $N_s$ samples of these circuits to solve {\sl XEB}, while typically classical algorithms either simulate these circuits noiselessly or 
approximately~\cite{markov2018quantum,villalonga2019flexible}. We do not rule out alternative classical algorithms that completely bypass circuit simulation, but do not consider this possibility here given the conjectured exponential classical cost of RCS.

It follows from the central limit theorem that $\sigma = N_s^{-1/2}$, i.e., that we must collect $N_s =O(\fid^{-2})$ samples from the QC, either from different random circuits or from the same random circuit, to ensure $\sigma \leq \fid$~\cite{supp} where in the simplest case we are assuming independent sampling. 

One might worry that a randomly generated circuit will be classically easy, e.g., because the distribution of hardness concentrates on easy cases. This is indeed the case for very small $m$~\cite{napp2019efficient}, or for circuits defined over planar graphs with only $O(\log(n))$ non-nearest-neighbor two-qubit gates~\cite{Geraci:2010gf}. However, standard conjectures invoked in quantum supremacy theory suggest that the distribution of hardness does concentrate on hard cases for large enough depth or sufficiently many non-nearest-neighbor two-qubit gates. The reason is as follows. Every probability $p_U(x) = |\bra{x}U\ket{0}|^2$ maps directly to the partition function of an Ising model at imaginary temperature~\cite{Lidar:04,Boixo:2016aa}. It is a strongly held conjecture that most partition functions are hard to approximate, which would imply (with Stockmeyer's theorem and anti-concentration, which has been proven for RCS~\cite{Hangleiter2018anticoncentration}) that most circuits are hard to sample. Going beyond conjectures, Refs.~\cite{Bouland:2018aa,Movassagh:2019aa} proved the hardness of exactly calculating  $p_U(x)$ in the average case. Hence, we proceed under the same set of assumptions as the original Google supremacy experiment~\cite{google}.

\section{Asymptotic classical scaling advantage above a threshold depth}%
\label{sec:SA}

We address several classical simulation algorithms in this and the following sections for RCS with an $n$-qubit and $m$-cycle quantum circuit, including the Schr\"odinger algorithm~\cite{ibm,markov2018quantum}, Schr\"odinger-Feynman algorithm~\cite{Aaronson:2016aa,Chen_2018,markov2018quantum} and tensor networks~\cite{Markov:tensor,boixo2017simulation,chen2018classical,villalonga2019flexible,Villalonga_2020,gray2020hyper,huang2020classical}. First, however, we address the scaling of a quantum computer for comparison. Given a time scaling of $T_\text{Q} \sim m/\fid^2$ (assuming parallel readout) for the QC to evaluate $1/\fid^2$ samples for cross-entropy benchmarking, we evaluate the scaling for samples using our fidelity model:
\begin{align}
\label{eq:T_Q}
	T_\text{Q} &\sim m 2^{\lambda m(3n-\sqrt{n}) + 2\gamma n}.
\end{align}

Out of the different classical simulation algorithms we consider, we begin with the Schr\"odinger algorithm (SA)~\cite{ibm,markov2018quantum} due to its favorable asymptotic scaling. Since it provides a full fidelity simulation, an optimal implementation of SA allows us to simulate only one randomly generated circuit
and then repeatedly sample from its resulting amplitudes to solve {\sl XEB}. From Eq.~(107) of Ref.~\cite{supp}, this is completed in time 
\begin{align}
\label{eq:T_SA}
	T_\text{SA} &\sim mn 2^n = T_\text{Q} n 2^{n(1- 2 \gamma) - \lambda m(3n-\sqrt{n})}.
\end{align}
For sufficiently small constants $\lambda$ and $\gamma$, {\sl XEB} can be classically solved exponentially faster in $m$ and $n$ using SA for any 
$m$ greater than a threshold value $m_{\text{th}}(n)$, corresponding to an asymptotic classical advantage for RCS for circuits deeper than
\begin{align}
	m_{\text{th}}(n) = \frac{n(1-2\gamma)+\log_2 n}{\lambda(3n-\sqrt{n})} \overset{n \to \infty}{\longrightarrow} \frac{1-2\gamma}{3\lambda} .
\end{align}

For the Google Sycamore device with $n=53$, this threshold occurs at $m \approx 87$. If the experimentally achieved values of $\lambda, \gamma$ may be sustained for larger devices, an advantage for SA is achieved for all $m \gtrsim 71$ as $n \to \infty$. As these are relatively shallow depths, this result may be compared to the algorithm of Bravyi, Gosset and Movassagh (BGM)~\cite{bravyi2019classical} for simulating the circuit in time $T_\text{BGM} \sim n 2^{O(m^2)}$, yielding a classical exponential speedup over quantum RCS for fixed $m$. Hence, a classical exponential advantage is achieved for RCS in all cases, above a certain width-dependent threshold circuit depth set by the quantum hardware's fidelity. However, classical hardware limitations constrain the experimental realization of such speedups, and thus we turn to the existence of a quantum runtime advantage at non-asymptotic widths and depths.

\begin{figure*}[t]
\centering
\subfigure[\ $2.8\times$ Sycamore error rate ($1\%$)]{\includegraphics[height=0.36\textwidth]{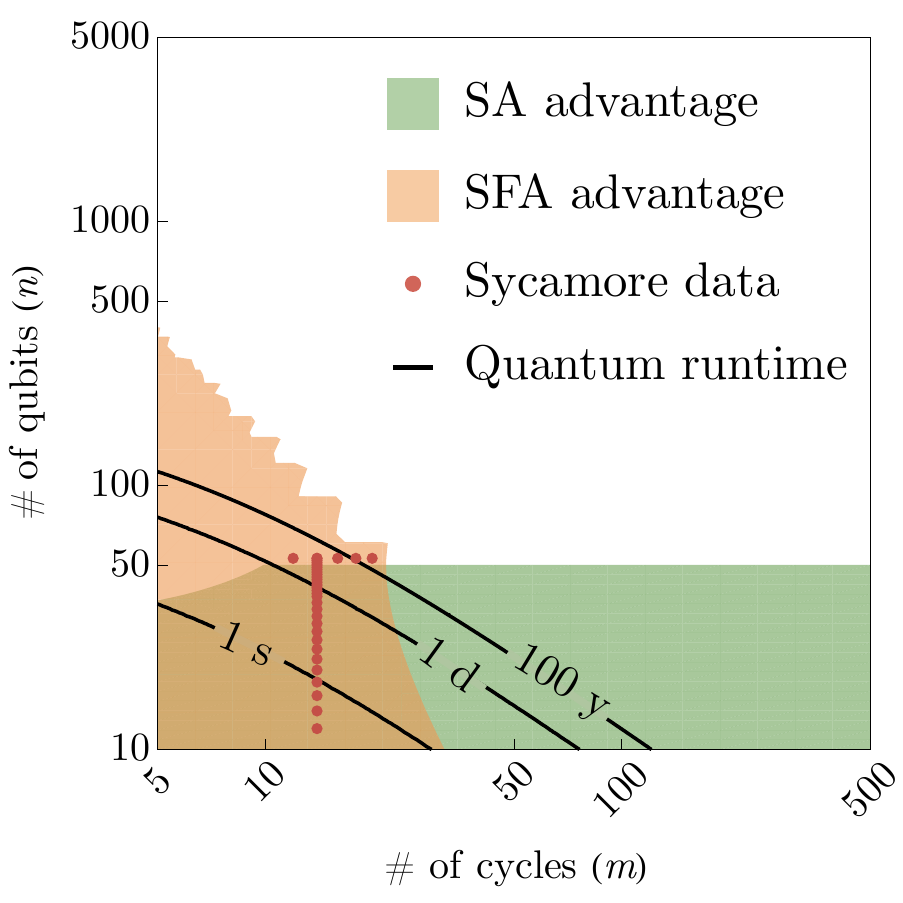}\label{fig:min(a)}}
\subfigure[\ Sycamore error rate ($0.36\%$)]{\includegraphics[height=0.36\textwidth]{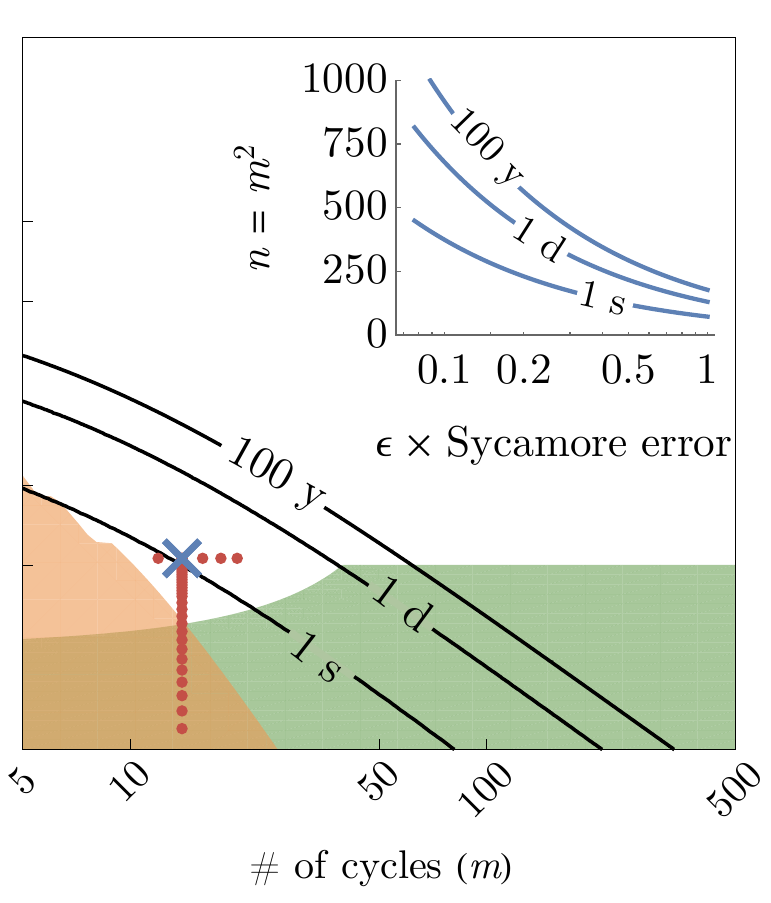}\label{fig:min(b)}}
\subfigure[\ $0.28\times$ Sycamore error rate ($0.1\%$)]{\includegraphics[height=0.36\textwidth]{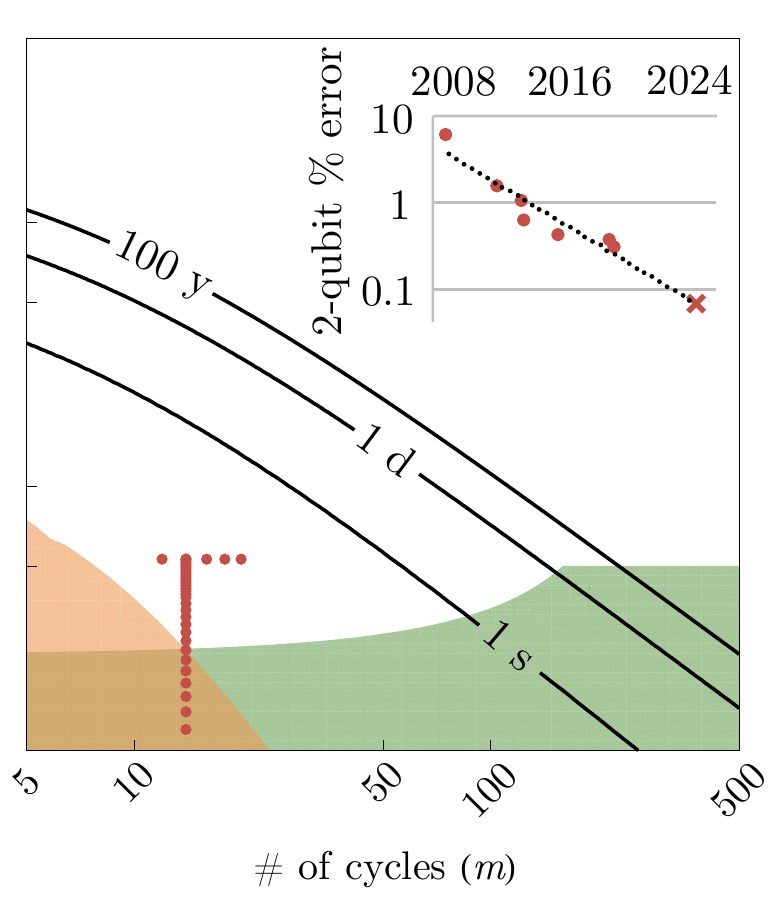}\label{fig:min(c)}}
\caption{(color online) Classical and quantum runtime advantage regimes at different error rates relative to the Sycamore error model fitted in Fig.~\ref{fig:qs}, with subfigure captions showing isolated two-qubit gate error rates. Black contours indicate quantum device runtimes; colored regions indicate where a classical runtime advantage is expected according to supercomputer memory and performance; red dots show the circuit width ($n$) and number of cycles ($m$) of each Sycamore experiment reported in~\cite{google}. Panel (a) shows that a mere $2.8\times$ factor increase in the error rate relative to the Sycamore chip would have required a quantum runtime of $100$ years to break even with SA and SFA. Panel (b) includes the extrapolated boundary of a TN runtime advantage at $(m,n)=(14,53)$, indicated by a cross. Panel (c) uses the extrapolated fidelity of a state-of-the-art NISQ device in 2025 (inset, error rates decay by a factor of $\sim0.77$ per year), and illustrates how even modest gains in error rates can significantly move the feasible quantum supremacy boundary. Runtimes are computed using Eqs.~\eqref{eq:T_Q},~\eqref{eq:T_SA}, and~\eqref{eq:T_SFA} with $p$ optimized within memory limitations, using performance and memory values reported in Ref.~\cite{supp}. Quantum {\sl XEB} runtime for circuits of depth $m=\sqrt{n}$ at error rates given by factors of the Sycamore error rate are shown in panel (b) inset. For reference, $\epsilon=1$ corresponds to an average isolated two-qubit gate error rate of 0.36\%~\cite{google}. All errors of the Sycamore device (single-qubit/two-qubit gates, readout errors) are scaled proportionally. Extrapolated error rates are given by an exponential regression over transmon device two-qubit gate errors~\cite{DiC09a,chow2012universal,chen2014qubit,barends_2014_superconductingquantum,sheldon2016procedure,google,kjaergaard2020quantum}.}
\label{fig:min}
\end{figure*}

\section{Limited-scale quantum runtime advantage}%
\label{sec:q-adv}

Due to limitations in random access memory, the Schr\"odinger algorithm is infeasible to run for a sufficiently large number of qubits, requiring storage of $2^n$ complex amplitudes. Similarly, the BGM algorithm has prohibitively large runtime with increasing depth since it scales as $2^{O(m^2)}$. In contrast, while achieving worse asymptotic performance than SA or BGM, the Schr\"odinger-Feynman algorithm (SFA)~\cite{Aaronson:2016aa,Chen_2018,markov2018quantum} and tensor networks (TN)~\cite{Markov:tensor,boixo2017simulation,chen2018classical,villalonga2019flexible,Villalonga_2020,gray2020hyper,huang2020classical} are more suitable to accommodate constraints of available classical hardware due to the use of Feynman paths and patching techniques that do not require keeping the entire quantum state in memory, as explained in more detail below. Both of these classical methods allow circuits to be simulated to partial fidelity $F$.

For SFA, we optimize the number of patches $p \ge 2$ and the number of paths simulated to satisfy {\sl XEB}.
We need to simulate  $2^{kpBm\sqrt{n}} F$ permutations or paths of Schmidt decompositions of cross-gates between patches~\cite{markov2018quantum,supp}.  After simulating each patch (at a time cost of $2^{n/p}$) we must compute the partial amplitudes of $1/F^2$ bitstrings (but at most $2^n$). Assuming that both simulation within patches and in between patches have similar runtime prefactors, the time scaling is
\begin{align}
\label{eq:T_SFA}
	T_\text{SFA} &= 2^{kpBm\sqrt{n}} F \left(p 2^{n/p} 
	+ \min\left(F^{-2}, 2^n \right)\right)
\end{align}
where $k=1$ for $p=2$ patches, $k=3/4$ for $p=4$ patches, and so on~\cite{Chen_2018,markov2018quantum,supp}, approximated as $k = \frac{1}{2} + \frac{1}{p}$. The constant $B=0.24$ is given by the grid layout of the Sycamore chip~\cite{supp}. Optimizing the runtime $T_\text{SFA}$ as a function of the simulation fidelity $F$ gives $F^{-2}=p 2^{n/p}$ for $n>\log_2(p)/(1-1/p)$.
In contrast to the SA memory usage of $2^n$ complex amplitudes, SFA with $p$ patches 
only requires $2 p 2^{n/p}$ complex amplitudes at the optimal simulation fidelity. Although increasing the number of patches reduces memory requirements, larger $p$ increases $T_\text{SFA}$ runtime as well. In practice, SFA runtimes may be improved by taking checkpoints during simulation~\cite{markov2018quantum}, but the leading order in runtime scaling is given by Eq.~\eqref{eq:T_SFA}.

To compare the expected runtime of a classical model to quantum hardware given our fidelity model, we define the total runtime of the computational task $R_x(n, m) = \tau_x T_x(n, m)$, with $x\in\{\text{Q},\text{SA},\text{SFA}\}$. The runtime constants 
$\tau_x$
are obtained from Ref.~\cite{supp}; for SFA this constant captures the fact that parallelization may reduce the runtime significantly. Since the total runtime is exponential in both $m$ and $n$ in all cases, and any scaling advantage is therefore due to a reduced exponent, we also define a more natural notion of an \emph{effective polynomial quantum speedup} $\alpha_{\text{C}}>0$ if for a classical method C (with, for our purposes, $\text{C} \in \{\text{SA},\text{SFA}\}$) the time scalings are related via $T_\text{C}(n, m) = T_\text{Q}(n, m)^{1+\alpha}$, i.e.,
\begin{align}
	\alpha_\text{C} &= \frac{\log T_\text{C}} { \log T_\text{Q}} - 1 .
\end{align}
A value of $\alpha_\text{C} > 0$ ($\alpha_\text{C} < 0$) implies a quantum (classical) advantage, capturing the relative reduction of the exponent in $T_{\text{Q}}(n, m)$ ($T_{\text{C}}(n, m)$). We compare the runtime advantage boundaries created by the effective polynomial speedup analysis and a direct runtime analysis. The result is shown in Fig.~\ref{fig:run}, both excluding runtime constants (shaded contours) and including runtime constants (dashed contours), which provides a region of quantum advantage indicated by the hatches. The Google Sycamore experiments are inside this boundary, which extends to a maximum depth and width of approximately $70$ cycles and $10^4$ qubits, respectively, beyond which classical algorithms dominate.

The results shown in Fig.~\ref{fig:run} do not account for memory constraints. To more accurately place bounds on a quantum runtime advantage, the memory limitations of classical hardware must be considered. An additional boundary is imposed by infeasible quantum runtimes, resulting in Fig.~\ref{fig:min}. Besides considering current quantum devices, we project superconducting qubit NISQ error rates and classical hardware to estimate the computational feasibility of {\sl XEB} in the near term. Error rates of two-qubit gates of superconducting qubit NISQ devices are taken from isolated measurements of transmons at a given year
~\cite{DiC09a,chow2012universal,chen2014qubit,barends_2014_superconductingquantum,sheldon2016procedure,google,kjaergaard2020quantum}
to determine an exponential fit shown in the inset of Fig.~\ref{fig:min(c)}, which is then applied to the fitted constants $\gamma$ and $\lambda$  in the error model of Eq.~\eqref{eq:F}. Although not exact, this provides an estimate of a reasonable range of error rates to consider on a 5-year timescale.

While SFA cuts the circuit into $p$ patches according to physical location, TN more generally finds optimal cuts in both circuit width and circuit depth and is found to be more efficient than SFA for deep enough circuits, including the Google experiment~\cite{gray2020hyper}. Because finding optimal TN contractions is \#P-hard for arbitrary circuits, we only include simulation results of TN performance at $n=53$ and $m=14$ from recent simulations~\cite{huang2020classical}, which improve on recent results in Ref.~\cite{gray2020hyper}. From Table~1 of Ref.~\cite{huang2020classical}, we scale a TN runtime of 88 seconds for $10^6$ samples at $F=1\%$ to $1/F^2 = 10^4$ samples, obtaining a TN runtime of 0.88 seconds.

\section{Implications for cross-entropy benchmarking and the significance of the quantum supremacy demonstration}%
\label{sec:implications}

We observe that the Google experiments have achieved a critical fidelity threshold to gain a runtime advantage over classical simulation. Had error rates been around $2.8\times$ larger than Sycamore's (corresponding to an increased isolated two-qubit gate error rate of $1\%$), no quantum advantage would have been achieved in cross-entropy benchmarking within 100-year quantum runtimes [Fig.~\ref{fig:min(a)}]. However, even at the fidelity achieved by the Google experiment, the quantum runtime advantage for {\sl XEB} stops at a few hundred qubits due to long quantum runtimes (Fig.~\ref{fig:min(b)}, which includes TN results).
Extrapolations suggest that even at achievable near-term fidelities below surface code thresholds, cross-entropy benchmarking will yield a runtime advantage up to at most a thousand qubits, beyond which quantum runtimes are computationally infeasible [Fig.~\ref{fig:min(c)}]. However, the regime of quantum advantage rapidly improves with lower error rates, underscoring the importance of achieving lower error rates for NISQ devices [Fig.~\ref{fig:min(b)}].

Given the constraints around a quantum runtime advantage for cross-entropy benchmarking in terms of fidelity and qubit count, we conclude that the onset of QEC will approximately coincide with the boundary of a feasible quantum advantage in random circuit sampling. Indeed, as shown in Fig.~\ref{fig:min(b)}, to perform {\sl XEB} with about $1,000$ qubits within reasonable time limits and at sufficient depth $m=\sqrt{n}$ to entangle any two qubits on a square lattice, error rates must improve by around an order of magnitude from Sycamore.

In the absence of QEC, circuits with $m \sim n$ require prohibitively many $1/\fid^2$ samples to run on a NISQ device at current Sycamore fidelity beyond around 50 qubits, leading to a Schr\"odinger algorithm advantage within supercomputer memory limitations. Outside the scope of {\sl XEB}, shallow circuits may allow for a quantum advantage~\cite{Bravyi:2017aa}: many near-term applications such as quantum simulation have $m \gtrsim n$~\cite{Babbush:2017aa}, although some important problems, such as the 2D Fermi-Hubbard problem, have a scaling $m \sim \sqrt n$~\cite{Jiang:2018aa}. Algorithms designed to provide a quantum advantage with shallow circuits may fare better without QEC given that both Figs.~\ref{fig:run} and~\ref{fig:min} show that increasing width is more beneficial than depth, at least for RCS. However, tensor network simulations --- although difficult to place precise bounds on their performance --- further tighten the bound on a quantum runtime advantage, particularly for shallow circuits with $m < \sqrt{n}$, which are upper-bounded by a time scaling of $T_\text{TN} \sim \fid 2^{O(m\sqrt{n})}$ for square lattices~\cite{Markov:tensor,Boixo:2016aa,boixo2017simulation}. 

As given by Eq.~\eqref{eq:F}, these results assume depolarization error with the additional simplification into independent cycle and qubit errors. The model provides a reasonable heuristic; the superexponential decay in fidelity for larger circuits visible in Fig.~\ref{fig:qs} largely occurs due to the increased proportion of two-qubit gates for circuits with a relatively smaller boundary, as seen by the better fit in Fig.~4 of Ref.~\cite{google} that considers gate-specific noise. If other sources of noise such as $1/f$ noise appear on longer timescales~\cite{Quintana:2017aa}, the quantum advantage region may be further constrained.

\section{Conclusion}%
\label{sec:conc}
Due to an asymptotic classical advantage for random circuit sampling at fixed quantum fidelity, we find that the projected quantum runtime advantages for the next five years in solving the {\sl XEB} problem underlying Google's quantum supremacy demonstration~\cite{google} are limited to the very early NISQ regime of $\sim 50$ qubits for circuits up to depth $\sim 200$ or up to $\sim 400$ qubits for shallower circuits. However, reducing the component error rate increases the quantum advantage regime exponentially, which underscores the importance of a continued emphasis on error rate reduction. We observe that while our work can be interpreted as placing a practical upper bound on circuit width and depth for which RCS-based quantum supremacy holds, a rigorous lower bound based on complexity theory conjectures ruling out all possibility of competitive classical simulation algorithms, both known and unknown, was presented in Ref.~\cite{alex2018qubits} for other supremacy proposals in the noiseless setting.

While RCS provides a purposeful milestone for measuring the progress of quantum devices, we have shown that the regime of quantum advantage for a standard approach to {\sl XEB} in the near term is upper-bounded by about a thousand qubits, rapidly approaching circuit sizes sufficient for early QEC. Given that fidelities comparable to those achieved in the Google experiment are close to establishing surface codes at a few hundred to a few thousand physical qubits~\cite{raussendorf2007fault, fowler2012surface, yoder2017surface,Bravyi:2018aa}, we anticipate that the disappearance of the quantum runtime advantage in RCS shown here approximately coincides with the onset of error-corrected quantum computing. Hence, while the point of comparison in {\sl XEB} and other metrics are either established through or motivated by classical simulation of random quantum circuits, this transition point into QEC motivates the adoption of problem-specific or QEC-specific metrics of progress for the field. By focusing on problems that do not require direct circuit simulation --- similar to the benchmarking of quantum annealers \textit{vs} state-of-the-art classical optimization algorithms~\cite{speedup,Albash:2017aa,PhysRevX.6.031015,Mandra:2017ab} --- we may obtain a more informative view of the usefulness of a quantum device in an architecture-agnostic manner. Within the NISQ regime, noise-resilient algorithms such as quantum many-body ground state preparation~\cite{kim2017noiseresilient},  
and tensor network contraction~\cite{kim2017robust}
might provide applications for quantum computers without fault tolerance~\cite{preskill2018quantum}, solving tasks 
that are classically approachable without direct simulation of quantum circuits. By comparing the quantum and classical runtimes and the resulting quality of the solution, an architecture-agnostic metric can be defined without reliance on a particular circuit simulation algorithm.

Many of the most appealing results for quantum computers 
are far more transformative in the presence of QEC. Given that fidelities comparable to those achieved in the Google experiment are close to establishing surface codes at a few hundred to a few thousand physical qubits~\cite{raussendorf2007fault, fowler2012surface, yoder2017surface,Bravyi:2018aa}, we expect that the disappearance of the quantum runtime advantage in RCS shown here will approximately coincide with the onset of error-corrected quantum computing. 
With this in mind we anticipate that given the limitations of the {\sl XEB} problem in maintaining a quantum advantage upon its intersection with early QEC circuit sizes, 
it will be natural to transition from the NISQ era to the QEC era along with metrics designed to capture performance under QEC, such as logical error rate.

\begin{acknowledgments}
The authors thank Tameem Albash, Fernando Brand\~ao, Elizabeth Crosson, and Maria Spiropulu for discussions, and Johnnie Gray for tensor network simulations of Sycamore circuits.
AZ and DAL acknowledge support from Caltech's Intelligent Quantum Networks and Technologies (INQNET) research program and by the DOE/HEP QuantISED program grant, Quantum Machine Learning and Quantum Computation Frameworks (QMLQCF) for HEP, award number DE-SC0019227. 
DAL further acknowledges support from the Oracle Corporation.
\end{acknowledgments}

\end{document}